\documentstyle[preprint,epsfig,aps]{revtex}
\begin{document}
\draft
\tightenlines
 \title{ On the close to threshold meson production \\
         in neutron-neutron collisions
       }
\author{P.~Moskal$^{1,2}$,
H.-H.~Adam$^3$,
A.~Budzanowski$^{4}$,
T.~G\"otz$^{2}$,
D.~Grzonka$^{2}$,
L.~Jarczyk$^{1}$,
A.~Khoukaz$^{3}$, 
K.~Kilian$^{2}$,
C.~Kolf$^{2}$,
P.~Kowina$^{2,5}$,
N.~Lang$^{3}$,
T.~Lister$^{3}$,
W.~Oelert$^{2}$,
C.~Quentmeier$^{3}$,
R.~Santo$^{3}$,
G.~Schepers$^{2}$,
T.~Sefzick$^{2}$,
M.~Siemaszko$^{5}$, 
J.~Smyrski$^{1}$, 
S.~Steltenkamp$^{3}$,
A.~Strza{\l}kowski$^{1}$,
P.~Winter$^{2}$,    
M.~Wolke$^{2}$,    
P.~W{\"u}stner$^{6}$,
W.~Zipper$^{5}$           
}
\address{$^1$        Institute of Physics, 
                   Jagellonian University, 
                   Cracow, Poland  }
\address{$^2$    Institut f{\"u}r Kernphysik, Forschungszentrum J\"{u}lich, 
                   Germany }
\address{$^3$    Institut f{\"u}r Kernphysik, Westf{\"a}lische Wilhelms--Universit{\"a}t,
                   M{\"u}nster, Germany}
\address{$^4$    Institute of Nuclear Physics, 
                   Cracow, Poland}
\address{$^5$    Institute of Physics, University of Silesia, 
                   Katowice, Poland }
\address{$^6$    Zentrallabor f{\"u}r Elektronik,  Forschungszentrum J\"{u}lich,
                   Germany }
\date{\today}
\maketitle
\begin{abstract}
     A method of measuring the close to threshold
     meson production in neutron-neutron collisions is described
     where the momenta of the colliding neutrons can be
     determined with the accuracy obtainable for 
     the proton-proton reaction. The technique
     is based on the double quasi-free $nn\rightarrow nn X^{0}$ reaction,
     where deuterons are used as a source of neutrons.
\end{abstract}
%
%
%
In the last decade close to threshold production of mesons
has attracted 
a lot of experimental and theoretical effort (see for instance~\cite{wilkin,machner}). 
 Experiments performed at the accelerators 
CELSIUS~\cite{bondarpl,caleneta,calenpdeta,calen_pneta,calen_higher},
COSY~\cite{smyrskipl,moskalpl,gem1,gem2,sewerinprl,bilgerpl,balewskipl,moskalprl},
IUCF~\cite{meyer90,meyer92,hardie,daehnick}, 
SATURNE~\cite{bergdolt,chiavassa,chiavassa94,hiboupl,disto_etap,disto_phiomega},
and
TRIUMF~\cite{hutcheon90,hutcheon91,korkmaz,bachman,duncan,hahn} 
delivered  precise data on the pseudoscalar ($\pi$,$\eta$,$\eta^{\prime}$,$K$)
and vector ($\omega$,$\phi$) meson production in 
proton-proton and proton-deu\-te\-ron collisions.
A secondary neutron beam with a spread in energy smaller than 1~MeV focussed
onto liquid hydrogen targets ($\sim$~10$^{23}$~atoms/cm$^{2}$) 
permitted also  precise investigations of the $\pi$ meson production 
in the neutron-proton reactions~\cite{hutcheon90,hutcheon91,bachman}.

Close to threshold meson production in proton-neutron
collisions were also investigated by means of a technique
based on a  quasi-free scattering
of the proton off the  neutron bound in the deuteron. 
Thin windowless internal deuterium cluster targets  
($\sim$~$10^{14}$~atoms/cm$^{2}$) make
a detection of an undisturbed spectator proton and 
a precise determination of the reacting neutron momentum --
and hence of the excess energy -- possible.\\
Pioneering experiments of the $\pi^{0}$ meson creation in the proton-neutron 
reaction
with the simultaneous
tagging of the spectator proton resulted in a resolution of the excess energy 
($\sigma$) equal to $\sigma$ = 1.8~MeV~\cite{bilgerspec}.
Similar studies including  the production  of heavier mesons will 
be continued at COSY~\cite{proposalankephiomega,proposalc11}.\\

\begin{figure}
\hspace{-0.1cm}
\resizebox{0.77\textwidth}{!}{%
\includegraphics{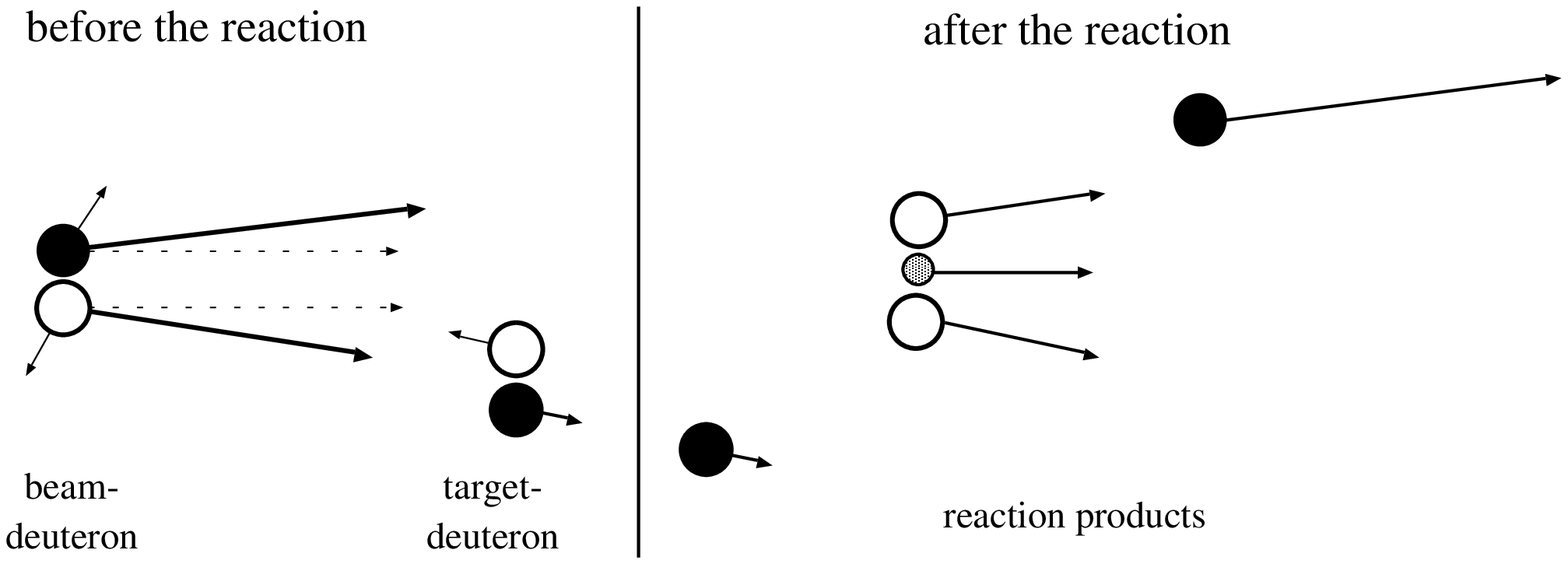}}
\vspace{1cm}
\caption{Schematic depiction of the double quasi-free 
         {\mbox{$ nn \rightarrow nnX$}} reaction.
         During the collisions of deuterons (left hand side of the figure, 
	 with the total momentum (solid arrow) resulting from the sum of the
	 beam momentum (dotted arrow) plus the Fermi momentum
	 (short arrow)) a double quasi-free neutron-neutron
         reaction may lead to the creation of mesons (small gray circle).
         The spectator protons~(black circles) leave the reaction region
         with their initial momentum plus the Fermi momentum, 
         which they possessed at the moment of the reaction.
         Neutrons are plotted as open circles.
         Due to the large relative momenta between spectators and the outgoing neutrons
         ($\sim$~1~GeV/c close to the threshold for the $\eta$ meson production)
         a distortion of the nnX system by the accompanied protons can be neglected.
        }
\label{dd_ppnneta}       
\end{figure}

Experimental investigations of the close to threshold
production in neutron-neutron collisions, however, have not yet been 
carried out. A realisation of such studies 
-- which are characterised by typical cross sections of $\le \mu$b --
with high quality neutron beams bombarding a deuterium target is not 
feasible due to the low neutron beam intensities forcing to use liquid or 
solid deuterium targets which make the precise determination of the momentum
of the spectator proton impossible.
In this contribution a unique possibility of the precise measurement of the 
close to threshold meson production in neutron-neutron collisions is 
pointed out. The technique is based on the double quasi-free interaction of 
neutrons originating from colliding deuterons as depicted in 
Figure~\ref{dd_ppnneta}.
Utilizing this method, a precision of $\sim$1~MeV can be obtained for 
determining the excess energy, since it depends only on the 
accuracy of the momentum or angle reconstruction for the registered
spectator protons. At present cooled deuteron beams -- available at 
the facilities CELSIUS, COSY, and IUCF -- give the possibility
of using this method for the studies of neutron-neutron scattering.
Moreover, the usage of a stored beam circulating through an internal
cluster target permits the study with high luminosities (~10$^{31}$cm$^{-2}$s$^{-1}$)
in spite of very low target densities.

In the double quasi-free interaction,
due to the small binding energy of the deuteron~($E_{B} = 2.2$~MeV),
the colliding neutrons may be approximately treated as 
free particles in the sense that the matrix element for quasi-free meson
production from bound neutrons is identical to that for the free
nn $\rightarrow $ NNX reaction at the same excess energy 
available in the NNX system.
The measurements at CELSIUS~\cite{calenpdeta,calen_pneta}
 and TRIUMF~\cite{duncan,hahn} have  proven 
that the offshellness of the reacting neutron
can be neglected and that the spectator proton influences the interaction
only in terms of the associated Fermi motion~\cite{duncan}.

\begin{figure}
\hspace{0.1cm}
\resizebox{0.77\textwidth}{!}{%
\includegraphics{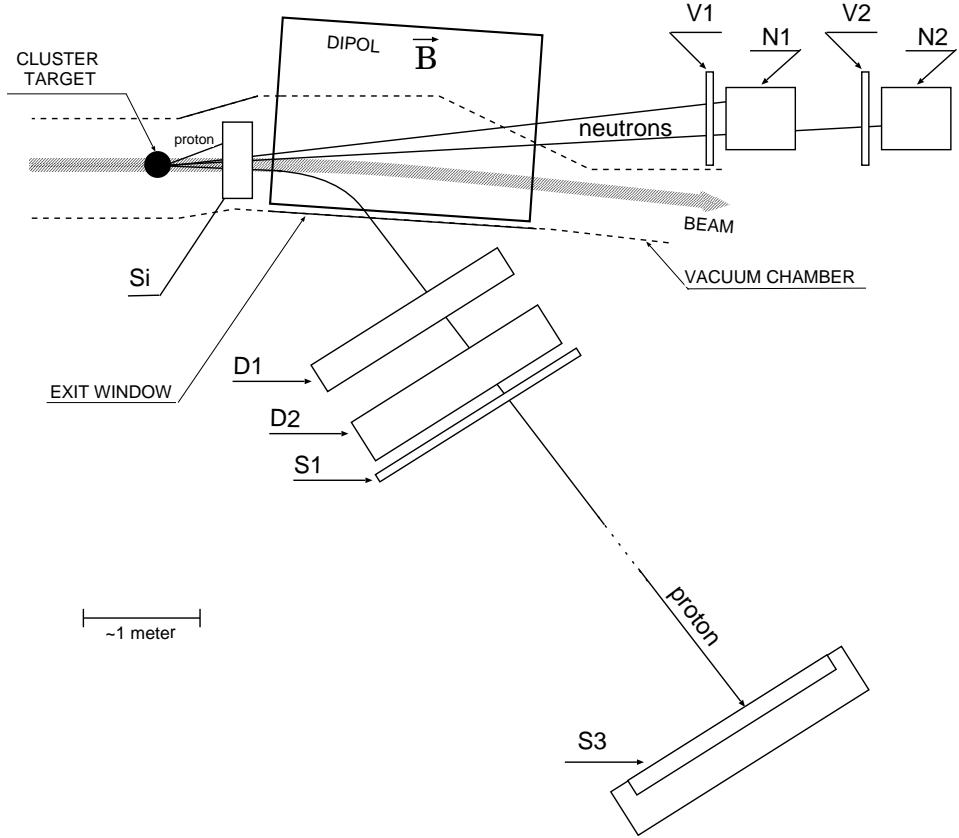}}
\vspace{1cm}
\caption{
    Schematic view of the extended COSY-11 detection setup~\protect\cite{c11nim}.
             Only detectors needed for the measurements of the 
	     $dd\rightarrow nn p_{sp}p_{sp} X$
             reactions are shown. \protect\\
             D1,D2 denote the drift chambers used for the track 
             reconstruction of the fast spectator proton;
             S1,S3 and V1,V2 are the scintillation detectors 
             used as  time-of-flight and veto counters, respectively,
             N1,N2  the neutron detectors, and 
	     Si~\protect\cite{bilgerspec} the silicon strip detectors.
 }
\label{detectors}       
\end{figure}

 The registration of both spectator protons will allow for a
precise determination of the excess energy. 
A possible internal target facility based on the COSY-11 setup~\cite{c11nim}
is presented in Figure~\ref{detectors}. The energy and the
emission angle of the "slow" spectator can 
be measured by an appropriately segmented silicon detector,
whereas the momentum of the "fast" spectator proton can be analysed
by the magnetic spectrometer. By means of the detection system
shown in Figure~\ref{detectors}, a resolution of the excess energy
of 2~MeV can be achieved for excess energies lower than 
30~MeV as demonstrated in reference~\cite{proposalc11}. 
The double quasi-free $nn\rightarrow nn X^{0}$ reaction
can be identified by the registration of both outgoing neutrons.
For example, in order to measure the production of the $\eta$ meson,
a four meter distance for the time of flight measurement would be enough
to obtain  8~MeV (FWHM) missing mass resolution~\cite{proposalc11}
with a calorimeter segmented 
by 10~cm~$\ast$~10~cm 
and providing a 0.5~ns~($\sigma$) time resolution, which was obtained in test
runs using a scintillator/lead sandwich type of detector.

The suggested meson production via a double 
quasi-free neutron--neutron reaction with precisions achievable
for the proton-proton and proton-neutron reactions,
opens the possibility of 
studying for example the charge symmetry breaking by comparing cross sections 
for the $pp\rightarrow pp\eta$ and $nn\rightarrow nn\eta$ reactions,
similarly to investigations performed via the $\pi$-deuteron reactions~\cite{tippens}. 
The Dalitz-plot analysis of the $ nn \rightarrow nn$ Meson would allow for the study
of the neutron-neutron and neutron-Meson~\cite{kudriavtsev} scattering lengths,
the first being still not well established~\cite{machleidt} and the second being unknown.
In principle when studying the meson production in proton-proton and in 
proton-neutron collisions one has access to all possible isospin combinations, 
which can be derived after the correction for the electromagnetic interaction.  
Exceptionally, close-to-threshold meson production via the
neutron-neutron scattering represents a pure $T=1$ isospin channel without 
accompanying coulomb interaction and consequently no need for its correction.
Investigations of neutron-neutron scattering allow also for the production 
of  $K^{+}K^{-}$ pairs in a system with only two charged particles in the final
state ($nn\rightarrow nnK^{+}K^{-}$), simplifying the theoretical
calculations drastically,
which in case of the $pp\rightarrow ppK^{+}K^{-}$ are not feasible due to the difficulty
of treating the  electromagnetic forces in the system of four charged particles~\cite{hanhart}.

At present the COSY synchrotron can accelerate deu\-te\-rons up to 3.5~GeV/c~\cite{prasuhnnim}
which, utilizing the Fermi momentum, allows for the $\pi$ and $\eta$ meson production
in the $nn\rightarrow nn X^{0}$ reaction. 
To investigate the neutron-neutron interaction
with the production of heavier mesons like $\omega$, $\eta^{\prime}$, or $\phi$,
a deuteron beam of $\sim 7$~GeV/c would be required.\\

We would like to thank V.~Baru, A.~Gasparyan, J.~Hai\-den\-bauer, 
Ch.~Hanhart, A.~Kudriavtsev, and 
C.~Wilkin for very helpful discussions.
%

%
%
\newpage

\end{document}